\documentstyle[preprint,aps,prd]{revtex}

\input{psfig}

\begin{document}

\draft
\preprint{}

\title{\bf Coefficient functions and open charm production in deep 
inelastic scattering} 

\author{A.V. Kisselev\thanks{Email address: kisselev@mx.ihep.su} \\
\small \em Instituto de F\'{\i}sica Gleb Wataghin, UNICAMP \\
\small \em 13083-970, Campinas, SP, Brasil \\
\small \em and \\
\small \em Institute for High Energy Physics, 142284 
Protvino, Russia\thanks{Permanent address}}
\date{}
 
\maketitle

\begin{abstract}
It is shown that the problem of double counting in open charm 
production in DIS can be solved by using the expression for DIS 
coefficient functions in terms of 2PI diagrams. \\ \\
PACS number(s): 13.60.Hb, 13.85.Ni, 14.65.Dw, 12.38.-t
\end{abstract} 
\vfill \eject

Open charm production in deep inelastic scattering (DIS) is a 
subject of interest from both experimental and theoretical point of view. 
Recent data from H1~\cite{H1} and ZEUS~\cite{ZEUS} collaborations have 
shown that charm quark contribution is an important component of DIS 
structure functions.

On of the most predictive methods of calculating open charm 
contribution to the structure functions is fixed order perturbative 
QCD (see \cite{Gluck}). Much efforts have been also done in order to 
formulate variable flavor number scheme (see 
\cite{Aivazis,Thorne,Martin,Kretzer}).

According to the factorization 
theorem~\cite{Radyushkin,Ellis,Curci,Collins}, the contribution of 
charm, $F_2^c$, can be represented as follows%
\footnote{Let us note that at very small values of $x$ one has to 
consider more general $k_{\bot}$--factorization \cite{Catani}}:
\begin{equation}
\frac{1}{x} F_2^c(Q^2,x) = \int_x^1 \frac{dz}{z} \left[ 
C^{c}(Q^2,\mu^2,z) q_c(\mu^2, \frac{x}{z}) + C^{g}(Q^2,\mu^2,z) 
g(\mu^2, \frac{x}{z}) \right]. 
\label{10}
\end{equation}
Here $C^{a}$ are the process--dependent coefficient functions ($a=c,g$), 
$q_c$ and $g$ being charm and gluon densities of the incoming hadron 
and $\mu$ is a factorization scale.
The gluon coefficient function, $C^{g}$, includes, in particular, 
photon--gluon fusion (PGF) contribution which dominates at low 
scales, $Q^2 < m_c^2$, where $m_c$ is a mass of charm quark. At high 
$Q^2$, as was noted in Refs.~\cite{Aivazis,Martin}, part of the PGF 
cross section is generated by the evolution of the charm contribution 
(the first term in Eq.~(\ref{10})).

To avoid double counting, one has to subtract this contribution. 
In Ref.~\cite{Aivazis} this subtraction has been taked into 
account in lowest order in $\alpha_s$ (in what follows a symbol 
$\otimes$ means a convolution in variable $z$):
\begin{equation}
F_2 = C_c^{(0)} \otimes q_c - C_c^{(0)} \otimes P_{cg} \otimes g +  
C_g^{(1)} \otimes g + \mbox{\rm O}(\alpha_s^2). \label{20}
\end{equation}

If $\mu \sim m_c$, we have an approximate cancellation of the first 
two terms in Eq.~(\ref{20}) and we arrive at the dominance of PGF 
mechanism. On the other hand, if $\mu \gg m_c$, the last two terms 
in Eq.~(\ref{20}) are almost cancell and we reproduce the QCD 
parton model in the leading order.

In fact, in Eq.~(\ref{20}) one should add another order $\alpha_s$ 
contribution of the form~\cite{Aivazis}:
\begin{equation}
C_c^{(1)} \otimes q_c - C_{c'}^{(0)} \otimes P_{c'c} \otimes q_c 
\label{21}
\end{equation}
which is numerically less important.

Let us note that the Wilson type coefficients in (\ref{10}) are
different from the coefficient functions $C_c$ and $C_g$ in (\ref{20})
and (\ref{21}). While the formers have no infrared or collinear 
singularities, the same is not true for the latters.

In Ref.~\cite{Martin} the modification of the gluon coefficient 
function has been proposed:
\begin{equation}
C_g \rightarrow C_g' = C_g^{PGF} - \Delta C_g. \label{30}
\end{equation}
By using the well--known PGF expression in the first order in 
$\alpha_s$ (see, for instance, \cite{Witten}), $\Delta C_g^{(1)}$ 
was calculated in \cite{Martin}. At $Q^2/m_c^2 \gg 1$ it looks like
\begin{equation}
\Delta C_g^{(1)} (Q^2,z) = \frac{\alpha_s}{2\pi} \left[ 
P_{cg}(z) \ln \frac{Q^2}{m_c^2} + z(1 - z) \right]. \label{31}
\end{equation}
The corresponding order $\alpha_s^2$ expression was also found (with 
the account of only large logarithmic terms)~\cite{Martin}:
\begin{eqnarray}
\Delta C_g^{(2)} &=&  C_c^{(0)} \otimes \left[ (\alpha_s \ln Q^2)^2 
(P_{cg}^{(0)} \otimes P_{gg}^{(0)} + P_{cc}^{(0)} \otimes 
P_{cg}^{(0)}) + \alpha_s^2 \ln Q^2 P_{cg}^{(1)} \right] 
\nonumber \\
&+&  C_c^{(1)} \otimes \alpha_s \ln Q^2 P_{cg}^{(0)} + 
\alpha_s C_g^{(1)} \otimes \alpha_s \ln Q^2 P_{gg}^{(0)}. 
\label{40}
\end{eqnarray}

Both procedures, (\ref{20}) and (\ref{30}), treat the problem of 
double counting {\em by hand} and they cannot be easily generalized 
to higher orders in $\alpha_s$. Let us note that expression 
(\ref{40}) was obtained with the account of only leading logarithmic
contributions in each subtracted term, and the 
factorization scale was assumed to be large, $\mu^2 \simeq Q^2$.

To overcome these shortcomings, let us start from a definition of 
coefficient functions in terms of two particle irreducible (2PI) 
amplitudes. In the following we assume that a corresponding object 
is 2PI in the direction of the iteration (that is in the $t$--channel).
We will work in an axial gauges ($n^{\mu} A_{\mu} = 0$, 
$n_{\mu}$ being a gauge vector) and  follow a scheme developed in 
Ref.~\cite{Ellis}.

Let $D$ being a matrix of formal parton distributions (below we 
define physical parton distribution functions $\tilde D$ 
(\ref{130})). It obeys a matrix equation
\begin{equation}
D = (I + \hat V + \hat V^2 + \cdots) \hat D = 
(I - \hat V)^{-1} \hat D, \label{60}
\end{equation}
where $\hat D$ is 2PI part of $D$, 2PI kernel $\hat V$ defines the 
evolution of $D$ and $I$ is a unit matrix. To simplify notations, 
we dropped a sum both in parton types and spins. Integration in 
internal momenta are also assumed in (\ref{60}).

It is known that in the axial gauge 2PI aplitudes have no 
singularities associated with a propagation of intermediate 
physical states~\cite{Ellis}. For instance, $\hat 
V(k_n, k_{n-1})$ in (\ref{60}) is finite at $k_{n-1}^2 
\rightarrow 0$, but it includes legs, corresponding to momentum 
$k_n$, and has a pole in $k_n^2$ as $k_n^2 \rightarrow 0$.  

Then DIS structure function has the form
\begin{equation}
F = \hat F + \hat A (I - \hat V)^{-1} \hat D \label{70}
\end{equation}
with $\hat F$ being the 2PI part of $F$, while $\hat A$ is the 
2PI part of the virtual photon--parton amplitude.

Following \cite{Ellis}, let us  introduce a projector operator 
onto physical helicity states ($d_{\mu \nu}$ is a tensor part of 
the gluon propagator)
\begin{equation}
P_h = \left\{
\begin{array}{cl}
{\displaystyle (\hat k)^{\alpha' \beta'} \left. \right] \left[ 
\right. (\frac{\hat n}{4kn})_{\alpha \beta}}, & \mbox{\rm 
for a quark line} \\
{\displaystyle \frac{1}{2} d^{\mu' \nu'} \left. \right] \left[ 
\right. (-g_{\mu \nu})}, & \mbox{\rm for a gluon line} 
\end{array}
\right. \label{80}
\end{equation}
and another operator which projects onto small virtualities of 
parton lines%
\footnote{Another possibility is to use an operator 
$P_{\varepsilon}$ which extracts poles in $\varepsilon$ in 
$\overline{\mbox{\rm MS}}$ renormalization scheme (see, for 
instance, \cite{Catani*}). We prefer to use the operator 
$P_{\mu}$ in order to have clear physical meaning for the 
scale $\mu$ (see below).}
\begin{equation}
P_{\mu} = \theta (\mu^2 - |k^2|). \label{90}
\end{equation}
As we will see, our result (\ref{154}) depend on a product of 
$P_h$ and $P_{\mu}$ (\ref{160}). But at intermediate steps  it 
is convenient to consider these operators separately.

By using matrix identities 
\begin{eqnarray}
(I - \hat V)^{-1} &=& [I - (I - P_h) \hat V)]^{-1} 
(I - P_h V')^{-1}, \nonumber \\
(I - P_h V')^{-1} &=& (I - \tilde V)^{-1}  
[I + P_{\mu} P_h \hat V (I - \hat V)^{-1}] \label{91}
\end{eqnarray}
with
\begin{equation}
V' =  \hat V [I - (I - P_h) \hat V]^{-1}, \qquad
\tilde V = (I - P_{\mu}) P_h V', \label{100}
\end{equation}
we can rewrite Eq.~(\ref{70}) in the form
\begin{eqnarray}
F &=& \left. \tilde A (I - \tilde V)^{-1} 
\right|_{k_{\bot}=k_-=0} 
\tilde D + \Delta \tilde A (I - \Delta \tilde V)^{-1} \tilde D 
\nonumber \\
&+& \left. \tilde A (I - \tilde V)^{-1} 
\right|_{k_{\bot}=k_-=0} 
\Delta \tilde V (I - \Delta \tilde V)^{-1} \tilde D 
+ \hat F, \label{110}
\end{eqnarray}
where
\begin{equation}
\tilde A = \hat A [I - (I - P_h) \hat V]^{-1} \label{111}
\end{equation}
and
\begin{eqnarray}
&& \Delta \tilde A = \left. \tilde A - \tilde A (q,k) 
\right|_{k_{\bot}=k_-=0}, \nonumber \\
&& \Delta \tilde V = \left. \tilde V - \tilde V (r,k) 
\right|_{k_{\bot}=k_-=0}. \label{120}
\end{eqnarray}
Here $k_-=(k^2 + k_{\bot}^2)/kn$.

In Eq.~(\ref{110}) we defined the quantity
\begin{equation}
\tilde D = [I + P_{\mu} P_h \hat V (I - \hat V)^{-1}] \hat D. 
\label{130}
\end{equation}

The quantity $\Delta \tilde A (I - \Delta \tilde V)^{-1}$ 
in the second term in Eq.~(\ref{110}) has no 
contributions of the type $\alpha_s^n (\ln Q^2)^k$. As far 
as one is interested in logarithms of $Q^2$, it is 
possible to omit this therm in the coefficiet function. 
The third term in Eq.~(\ref{110}), $\Delta \tilde V(I - 
\Delta \tilde V)^{-1} \tilde D$, is of the oder of 
$\Lambda^2/\mu^2$, where $\Lambda$ is a tipical hadron scale. 
The non--perturbative scale $\Lambda$ arises as a result of 
integration of the kernel $\hat V$  with $\hat D$ (which 
describes initial parton distributions inside the nucleon).
The quantity $\hat F$ in (\ref{110}) is related to higher 
twist effects and it is suppressed by a factor 
$\Lambda^2/Q^2$.

So, at high $Q^2$ and $\mu^2 \gg \Lambda^2$ we get:
\begin{equation}
F = C \tilde D, \label{140}
\end{equation}
where
\begin{equation}
\left. C = \tilde A (I - \tilde V)^{-1} (q,k) 
\right|_{k_{\bot}=k_-=0} \label{150}
\end{equation}
with $\tilde V$ and $\tilde A$ defined above (see (\ref{100}), 
(\ref{111})).

Let us note that the quantity $\tilde D$ (\ref{130}) is of 
the sum of the terms wich looks like the following (after an 
integration in transverse components of momenta $k_i$):
\begin{eqnarray}
&& \int^{\mu^2} dk_n^2 \int^{k_n^2} dk_{n-1}^2 \int 
\frac{dz_{n-1}}{z_{n-1}} \hat V(z_{n-1}, 
\frac{k_{n-1}^2}{k_n^2}) \int^{k_{n-1}^2} dk_{n-2}^2  
\nonumber \\ 
&& \int \frac{dz_{n-2}}{z_{n-2}} \hat V(z_{n-2}, 
\frac{k_{n-1}^2}{k_{n-2}^2}) \ldots  \int^{k_1^2} dk_0^2 
\int \frac{dz_0}{z_0}\hat V(z_0, \frac{k_0^2}{k_1^2} ) 
\hat D(z_0,k_0^2, p^2), \label{151}
\end{eqnarray}
where $z= k_{n+1} n/k_n n$.
On the other hand, from an analogous formula in \cite{Ellis} 
we get the terms of the type
\begin{eqnarray}
&& \int^{\mu^2} dk_n^2 \int^{k_n^2} dk_{n-1}^2 \int 
\frac{dz_{n-1}}{z_{n-1}} U(z_{n-1}, 
\frac{k_{n-1}^2}{k_n^2}) \int^{k_{n-1}^2} dk_{n-2}^2  
\nonumber \\ 
&& \int \frac{dz_{n-2}}{z_{n-2}} U(z_{n-2}, 
\frac{k_{n-1}^2}{k_{n-2}^2}) \ldots \int \frac{dz_0}{z_0}
U(z_0, \frac{k_0^2}{k_1^2} ), \label{152}
\end{eqnarray}
where
\begin{equation}
U = P_{\mu} P_h \hat V [I - (I - P_h) \hat V]^{-1}. 
\label{153}
\end{equation}

As one can easily see from (\ref{130}), (\ref{151}), 
$\tilde D$ has the meaning of the distribution of partons 
whose virtualities vary ut to $\mu^2$. That is why we can 
consider $C$ in Eq.~(\ref{140}) to be a coefficient 
function. It can be represented in another (equivalent) 
form:
\begin{equation}
\left. C = \hat A [I - (I - K) \hat V]^{-1} (q,k) 
\right|_{k_{\bot}=k_-=0}. \label{154}
\end{equation}
Here 
\begin{equation}
K = P_{\mu} P_h. \label{160}
\end{equation}
So, the operator $(I - K) = (I - P_h) + (I - P_{\mu}) 
P_h$ in Eq.~(\ref{154}) projects onto states without 
any collinear singularities. It acts on the full 
expression on the right~\cite{Curci}:
\begin{equation}
(I - (I - K) \hat V)^{-1} = I + (I - K) \hat V 
+ (I - K) (\hat V (I - K) \hat V) + \ldots . 
\label{161}
\end{equation}
The expression for $C_g^{(1)}$, calculated with the use 
of formula (\ref{154}), coincides with $\alpha_s$ order 
result from Ref.~\cite{Mathews}.

Starting from Eq.~(\ref{150}) or (\ref{154}), we can 
represent DIS coefficient function $C$ in the form 
analogous to (\ref{30}):
\begin{equation}
\left. C = \left( C^{PhP} - \Delta C \right) (q,k) 
\right|_{k_{\bot}=k_-=0}, 
\label{162}
\end{equation}
where 
\begin{equation}
C^{PhP} = \hat A (I - \hat V)^{-1}  \label{163}
\end{equation}
is a "naive" expression for virtual photon--parton 
coefficient function that does not take into account 
the evolution included in parton distributions 
$\tilde D$ (\ref{130}). In particular, if we consider a 
parton to be a gluon, we have $C_g^{PhP} = C^{PGF}$.

From Eqs.~ (\ref{162}), (\ref{154}) and (\ref{163}) we 
get the following expression for $\Delta C$:
\begin{equation}
\Delta C = \tilde C K \hat V (I - \hat V)^{-1}, 
\label{170}
\end{equation}
where $\tilde C(q,k)$ is given by formula (\ref{154}) 
but without imposing a condition $k_{\bot}=k_-=0$.
In deriving (\ref{170}) we used a matrix identity
\begin{equation}
(I - \hat V)^{-1} - [I - (I - K) \hat V]^{-1} = 
[I - (I - K) \hat V]^{-1} K \hat V (I - \hat V)^{-1}. 
\label {171}
\end{equation}

Let us note that both $C^{PhP}(q,k)$ (\ref{163}) and 
$\Delta C(q,k)$(\ref{170}) contain, in general, 
singularities at $k^2=0$, while their difference, 
$C(q,k)$ (\ref{154}), does not.

The formula (\ref{170}) enables us to calculate 
$\Delta C_g$ in any fixed order in $\alpha_s$. Let the 
coefficient function $C$ and the kernel $\hat V$ have, 
respectively, the expansions:
\begin{equation} 
C = C^{(0)} + C^{(1)} + \ldots \label{172}
\end{equation}
and
\begin{equation} 
\hat V = \hat V^{(0)} + \hat V^{(1)} + \ldots. 
\label{173}
\end{equation}
Then we get:
\begin{equation}
\Delta C_g^{(1)} = \tilde C_c^{(0)} K \hat V_{cg}^{(1)} 
\label{180}
\end{equation}
and
\begin{eqnarray}
\Delta C_g^{(2)} &=& \tilde C_c^{(0)} K 
[\hat V_{cg}^{(1)} \hat V_{gg}^{(1)} +  \hat V_{cc}^{(1)} 
\hat V_{cg}^{(1)}] + \tilde C_c^{(0)} K \hat V_{cg}^{(2)} 
\nonumber \\
&+& \tilde C_c^{(1)} K \hat V_{cg}^{(1)} + \tilde 
C_g^{(1)} K \hat V_{gg}^{(1)}. \label{190}
\end{eqnarray}

The kernels $\hat V$ can be related with parton 
splitting functions ($a, b = q, g$)~\cite{Catani*}:
\begin{equation}
\hat V_{ab}^{(n)} (z, r^2, k^2) = \frac{1}{r^2} 
\frac{\alpha_s}{2\pi} \hat P_{ab}^{(n-1)} \left(z, 
\frac{k^2}{r^2} \right), \label{200}
\end{equation}
where off--shell splitting function has the form:
\begin{equation}
\hat P_{ab} = \left( \frac{\alpha_s}{2\pi} \right) 
\hat P_{ab}^{(0)} + \left( \frac{\alpha_s}{2\pi} 
\right)^2 \hat P_{ab}^{(1)} + \ldots. \label{201}
\end{equation}
In particular, in the leading logarithmic approximation, 
we have:
\begin{equation}
\hat V_{ab}^{(1)} (z, r^2, k^2 = 0) = \frac{1}{r^2} 
\frac{\alpha_s}{2\pi} P_{ab}^{(0)}(z), \label{210}
\end{equation}
where $P_{ab}^{(0)}(z)$ is a leading order Altarelli--Parisi 
splitting function.

If we put $\mu^2 \simeq Q^2$ in Eq.~(\ref{190}) and save 
$(\alpha_s \ln Q^2)^2$ and $\alpha_s^2 \ln Q^2$ 
contributions, all the terms in expression 
(\ref{40}) for $\Delta C_g$ can be  reproduced (taking 
into account slight difference between our definitions of 
$C^{(n)}$ and  $\Delta C^{(n)}$ and those from 
Ref.~\cite{Martin}). Indeed, from (\ref{200}) we conclude 
that $C_c^{(1)} K \hat V_{cg}^{(1)} \simeq C_c^{(1)} \otimes 
\alpha_s \ln Q^2 P_{cg}^{(1)}$, etc., where coefficient 
functions $C_a^{(n)} = C_a^{(n)}(\alpha_s(Q^2))$ have no 
logarithms of $Q^2$. 

However, exact formula (\ref{190}) results in additional 
contributions which are absent in (\ref{40}). In particular, 
due to power corrections in $\hat V_{ab}^{(1)}$ (see 
(\ref{200})) the term $C_c^{(0)} K [\hat V_{cg}^{(1)} 
\hat V_{gg}^{(1)} +  \hat V_{cc}^{(1)} \hat V_{cg}^{(1)}]$ 
contains a non--leading contribution ($\alpha_s^2 \ln Q^2$) 
in addition to a leading one ($(\alpha_s \ln Q^2)^2$).
 
Let us note that expression (\ref{170}) and, consequently, 
(\ref{180}) and (\ref{190}) do not have factorize forms as 
right hand sides of Eqs.~(\ref{20}), (\ref{40}) do. Namely, 
the integration of $\tilde C(q,l)$ or $\tilde C^{(n)}(q,l)$ 
in the momenta $l$ with the right part of the corresponding 
expression should be done. In particular, due to this 
integration in equation (\ref{180}) we get a constant term in 
$\Delta C_c^{(1)}(q,k)$ in addition to large logarithmic term.  
It depends both on a ratio $\mu^2/Q^2$ and $\mu_0^2/Q^2$, 
where $\mu_0^2 = - k^2$ (simultaneously, we have to put 
$k^2 = - \mu_0^2$ in $C^{PhP}(q,k)$). At $\mu^2 \simeq Q^2 \gg 
m_c^2$ and $k^2 = 0$ we get the constant term which is different 
from $z(1 - z)$ in (\ref{31}).  

Fortunately, we are not forced to deal with the quantity 
$\Delta C$ (\ref{170}) as we have derived the formula 
(\ref{154}) which enables us to calculate $C_g$ and $C_c$ in 
any order in strong coupling without getting double counting in 
the coefficient functions.

In the present paper we have studied the structure function $F_2$.
However, our formulas (\ref{154}), (\ref{170}) may be also applied to 
the longitudinal deep inelastic structure function $F_L$.

In conclusion let us note that the problem of double counting 
should also exist for Wilson coefficients for light quarks, $C^{q}$, 
considered in higher orders in $\alpha_s$. In such a case 
the double counting means that one and the same term is accounted 
for both in the coefficient function $C^{q}$ and in a distribution 
function $\tilde D_q$. Formula (\ref{150}) ((\ref{154})) enables 
one to separate diagrams, describing $C^{q}$, from those included 
in the evolution of the quark distribution $\tilde D_q$. For instance, 
in the lowest order in $\alpha_s$ we should get a term analogous to 
(\ref{30}).
 
I would like to thank the Brazilian govemental agency FAPESP 
for financial support.
\vfill \eject

\end{document}